\documentstyle[preprint,aps,epsfig]{revtex}

\begin{document}

\title{Continuous-wave Doppler-cooling of hydrogen atoms with two-photon transitions}
\author{V{\'e}ronique Zehnl{\'e} and Jean Claude Garreau}
\address{Laboratoire de Physique des Lasers, Atomes et Mol{\'e}cules and
Centre d'Etudes et de Recherches Laser et Applications\\
Universit\'{e} des Sciences et Technologies de Lille \\
F-59655 Villeneuve d'Ascq Cedex, France\\
}
\maketitle

\begin{abstract}
We propose and analyze the possibility of performing {\em two-photon}
continuous-wave Doppler-cooling of hydrogen atoms using the $1S-2S$
transition. ``Quenching" of the $2S$ level (by coupling with
the $2P$ state) is used to increase the
cycling frequency, and to control the equilibrium temperature.
Theoretical and numerical studies of the heating
effect due to Doppler-free two-photon 
transitions evidence an increase of the temperature by a factor of
two. The equilibrium temperature decreases with the
{\em effective} (quenching dependent) width of the excited state
and can thus be adjusted up to values close to the recoil temperature.
\end{abstract}

%]

Laser cooling of neutral atoms has been a most active research field for
many years now, producing a great deal of new physics. Still, the hydrogen
atom, whose ``simple" structure has lead to fundamental steps in the
understanding of quantum mechanics, has not yet been laser-cooled.
The recent experimental demonstration of the Bose-Einstein condensation
of H adds even more interest on laser cooling of hydrogen \cite{ref:BECH}.
One of
the main difficulties encountered in doing so is that all transitions
starting from the ground state of H fall in the vacuum ultraviolet (VUV) range
(121 nm for the $1S-2P$ transition), a spectral domain where
coherent radiation is difficult to generate. In 1993, M. Allegrini and E.
Arimondo have suggested the laser cooling of hydrogen by two-photon $\pi$
pulses on the $1S-3S$ transition (wavelength of 200 nm for two-photon
transitions) {\cite{ref:Maria&Ennio}. Since then, methods for generation of
CW, VUV, laser radiation have considerably improved, and
have been extensively used in metrological experiments \cite{ref:1S2S}. This
technical progress allows one to realistically envisage the two-photon Doppler
cooling (TPDC) of hydrogen in the {\em continuous wave} regime, in
particular for the $1S-2S$ two-photon transition. }

Laser cooling relies on the ability of the atom to perform a great number of
fluorescence cycles in which momentum is exchanged with the radiation
field. It is well known that $2S$ is a long-lived metastable
state, with a lifetime approaching one second. From this point of view, the $%
1S-2S$ two-photon transition is not suitable for cooling. On the other hand,
the minimum temperature achieved via Doppler cooling is proportional to the
linewidth of the excited level involved on the process \cite{ref:Houches},
a result that will be shown to be also valid for TPDC. From this
point of view $2S$ is an interesting state.

In order to conciliate these antagonistic properties of the $1S-2P$
transition, we consider in the present work the
possibility of using the ``quenching" \cite{note:quenching} of the
$2S$ state to
control the cycling frequency of the TPDC process. For the sake of
simplicity, we work with a one-dimensional model. We write rate
equations describing TPDC on the $1S- 2S$ transition in presence of
quenching. The quenching ratio is considered as a free parameter,
allowing control of the equilibrium temperature. The
cooling method is then in principle limited only by photon recoil effects.

We also develop analytical approaches to the problem. A Fokker-Planck
equation is derived, describing the dynamics of the process for temperatures
well above the recoil temperature $T_r$ (corresponding to the kinetic
energy acquired by an atom in emitting a photon).
A numerical analysis of the dynamics of the cooling
process completes our study.

\begin{figure}[tbp]
  \begin{center}
    \epsfig{figure=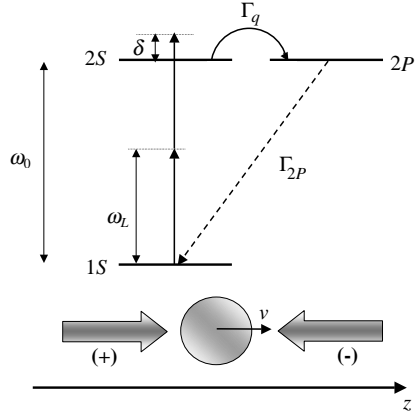,width=8cm,clip=}
    \end{center}
\caption{Hydrogen levels involved in the two-photon Doppler cooling in
  presence of quenching.}
\label{fig:Levels}
\end{figure}

Let us consider a hydrogen atom of mass $M$ and velocity $v$ parallel to the 
$z$-axis (Fig. \ref{fig:Levels}) interacting with two counterpropagating
waves of angular frequency $\omega_L$ with $2 \omega_L=\omega_0 + \delta$,
where $\omega_0/2\pi = 2.5 \times 10^{14}$ Hz is the frequency
corresponding to the transition $1S
\rightarrow 2S$, and also define the quantity $k \simeq 2
k_L=2\omega_L/c$.
The shift of velocity corresponding to the absorption of two-photons in
the {\em same} laser wave is $\Delta = \hbar k/M = 3.1$ m/s.
We will neglect the frequency
separation between $2S$ and $2P$ states (the Lamb shift -- which
is of order of 1.04 GHz) and
consider that the one-photon spontaneous desexcitation from the $2P$ states
also shifts the atomic velocity of $\Delta$ randomly in the $+z$ or $-z$
direction. Note that $T_r = M \Delta ^2/k_B \approx
1.2$ mK for the considered transition ($k_B$ is the Boltzmann
constant).
We neglect the photo-ionization process connecting the excited states to
the continuum. This is justified by the $1/E$ decreasing of the
continuum
density of states as a function of their energy $E$ and by the fact that a
monochromatic laser couples the excited levels only to a very
small range of continuum levels.

The atom is subjected to a controllable quenching process that couples the $%
2S$ state to the $2P$ state (linewidth $\Gamma _{2P}=6.3\times 10^{8}s^{-1}$%
). The adjustable quenching rate is $\Gamma _{q}$. Four two-photon
absorption process are allowed: {\it i}) absorption of two photons from the $%
+z$-propagating wave (named wave ``$+$'' in what follows), with a rate $%
\Gamma _{1}$ and corresponding to the a total atomic velocity shift of $%
+\Delta $; {\it ii}) absorption of two photons from the $-z$-propagating
wave (wave ``$-$''), with a rate $\Gamma _{-1}$ and atomic velocity shift of 
$-\Delta $; {\it iii}) the absorption of a photon in the wave ``$+$''
followed by the absorption of a photon in the wave ``$-$'', with no velocity
shift and {\it iv}) the absorption of a photon in the wave ``$-$'' followed
by the absorption of a photon in the wave ``$+$'',
with no velocity shift. The
two latter process are indistinguishable, and the only relevant transition
rate is that obtained by squaring the sum of the {\em amplitudes} of these
process (called $\Gamma _{0}$). Also, these process are
``Doppler-free'' (DF) as they are insensitive to the atomic velocity (to the
first order in $v/c$) and do not shift the atomic velocity. Thus, they cannot
contribute to the cooling process. As atoms excited by the DF process must
eventually spontaneously decay to the ground state, this process {\em heats}
the atoms. In the limit of low velocities, the transition amplitude
for each of the four processes is the same. One thus expects the DF
transitions to increase the equilibrium temperature by a factor
of two.

We can easily account for the presence of the quenching by
introducing an {\it effective} linewidth of the excited level
(which, due to the quenching process, is a mixing of the $2S$ and $2P$
levels) given by 
\begin{equation}
\Gamma _{e}=\Gamma _{2P}\text{ }{\frac{\Gamma _{q}}{\Gamma _{q}+\Gamma
_{2P}}}=g\Gamma _{2P}  \label{eq:GammeEff}
\end{equation}
with $g \equiv \Gamma _{q}/(\Gamma _{q}+\Gamma _{2P})$. This
approximation is true
as far as the quenching ratio is much greater than the width of the $2S$
state (note that this range is very large, as the width of the $2S$ state is
about $10^{-8}$ times that of the $2P$ state).

The two-photon transition rates \cite{ref:twophoton} are given by: 
\begin{equation}
\Gamma _{n}=\Gamma _{2P}{\frac{g}{2}}\text{ }{\frac{(1+3\delta _{n0})\bar{I}%
    ^{2}}{(\bar{\delta}-nKV)^{2}+g^{2}/4}}
\label{eq:twophotontransrates}
\end{equation}
where $n=\{-1,0,1\}$ describes, respectively, the absorption from the ``$-$%
'' wave, DF transitions, and the absorption from the ``$+$'' wave. $\bar{I}%
\equiv I/I_{s}$ where $I_{s}$ is the two-photon saturation intensity, $\bar{%
\delta}$ is the two-photon detuning divided by $\Gamma _{2P}$, $K\equiv
k\Delta /\Gamma _{2P} = 0.26$ and $V\equiv v/\Delta $.

The rate equations describing the evolution of the velocity distribution $%
n(V,t)$ and $n^{*}(V,t)$ for, respectively, atoms in the ground and in the
excited level are thus 
\begin{mathletters}
\begin{equation}
{\frac{\partial n(V,t)}{\partial t}}=-\left[ \Gamma _{-1}(V)+\Gamma
_{0}+\Gamma _{1}(V)\right] n(V,t)+{\frac{\Gamma _{e}}{2}}\left[
n^{*}(V-1)+n^{*}(V+1)\right]   \label{eq:rate_n}
\end{equation}
\begin{equation}
{\frac{\partial n^{*}(V,t)}{\partial t}}=\Gamma _{-1}(V-1)n(V-1,t)+\Gamma
_{0}n(V,t)+\Gamma _{1}(V+1)n(V+1,t)-\Gamma _{e}n^{*}(V,t)\text{.}
\label{eq:rate_nstar}
\end{equation}
The deduction of the above equations is quite straightforward
(cf. Fig \ref{fig:Levels}).
The first term in the right-hand side of Eq. (\ref{eq:rate_n})
describes the depopulation of the ground-state velocity class $V$
by two-photon transitions, whereas the second term describes the
repopulation of the same velocity class by spontaneous decay from the
excited level. In the same way, the three first terms in the right-hand side
of Eq. (\ref{eq:rate_nstar}) describe the repopulation of the excited
state velocity class $V$ by two-photon transition, and the last term
the depopulation of this velocity class by spontaneous transitions.
For each term, we took into account the velocity shift
($V \rightarrow V \pm 1$) associated with each transition and
supposed that spontaneous emission is symmetric under spatial inversion.

For moderate laser intensities, one can adiabatically eliminate the
population of excited level. This is valid far from
the saturation of the two-photon transitions and reduces the
Eqs. (3a-3b) to one equation describing
the evolution of the ground-state population:

\end{mathletters}
\begin{eqnarray}
{\frac{dn(V,t)}{dt}} &=&-\left[ \Gamma _{0}+\frac{\Gamma _{-1}(V)}{2}+
{\frac{\Gamma _{1}(V)}{2}}\right] n(V,t)+  \nonumber \\
&&{\frac{1}{2}} \{ \Gamma _{0}\left[ n(V-1,t)+n(V+1,t)\right] +\Gamma
_{-1}(V-2)n(V-2,t)+\Gamma _{1}(V+2)n(V+2,t) \}   
\label{eq:elim} 
\end{eqnarray}

Eq.(\ref{eq:elim}) is in fact a set of linear ordinary differential
equations coupling the populations of velocity classes separated by
an integer: $V,V \pm 1,V \pm 2,\cdots$. This discretization exists only 
in the 1-D approach considered here, but it does not significantly affect
the conclusions of our study, while greatly simplifying the numerical
approach.

Eqs.(\ref{eq:elim}) can be recast as $d{\bf n}/dt=C{\bf n},$ where $C$
is a square matrix and ${\bf n}$ is the vector 
$(\cdots n(-i,t),\cdots n(0,t),n(1,t),\cdots)$. Numerically, the 
equilibrium distribution is obtained in a simple
way as the eigenvector ${\bf n_{eq}}$ of $C$ with zero eigenvalue. In this way,
the asymptotic temperature is obtained as : 
\begin{equation}
\frac{T}{T_{r}}=\left\langle V^{2}\right\rangle = \frac{%
\sum\limits_{i=-\infty }^{\infty }i^{2}n_{eq}(i)}{\sum\limits_{i=-\infty
}^{\infty }n_{eq}(i)}
\end{equation}

\begin{figure}[tbp]
  \epsfig{figure=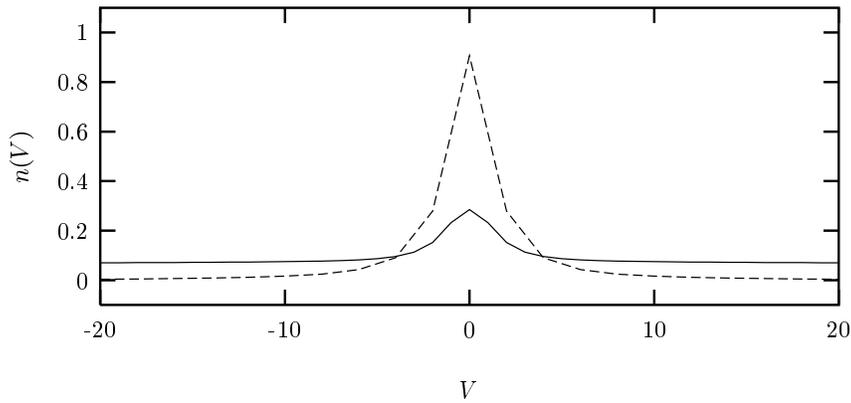,width=15cm}
  \caption{Numerically calculated velocity distributions
    with $\bar{\delta}=-0.25$ and $g=1/3$.
The dotted curve corresponds to the distribution obtained by
suppressing Doppler-free transitions (cf. text). Typically,
the distribution exhibits two structures: a broad background due to the
atoms heat by Doppler-free transitions and a sharp peak of cold atoms. }
\label{fig:distrs}
\end{figure}

Fig.\, \ref{fig:distrs} shows the equilibrium distribution
obtained by numerical simulation for $\bar{ \delta} = -0.25$ and
$g = 1/3$. The dotted
curve corresponds to the distribution obtained by artificially
suppressing DF transitions (i.e., by setting $\Gamma _{0}=0$).
As we pointed out earlier, the DF transitions lead to a heating effect.
Doppler cooling is efficient mainly for
atoms distributed on a range of $g/K$ around the velocity $V=\pm |%
\bar{\delta }|/K$ \cite{ref:Houches} whereas Doppler-free transitions
are independent of the velocity; all velocity classes are thus
are affected by the heating. As a consequence, DF transitions induce a
deformation of the velocity profile, specially for small values of $g$
and $\bar{ \delta }$,
superimposing a sharp peak of cold atoms on a wide background of ``hot''
atoms. In what follows, all numerically calculated-temperatures
are deduced form the width of the thin peak of cold atoms.

Eqs. (\ref{eq:rate_n}) and (\ref{eq:rate_nstar}) or
Eq. (\ref{eq:elim}) have no exact solution. However, using some
reasonable hypothesis, it is possible to develop analytical approaches.
The most usual of these approaches is to derive
from the above equations a Fokker-Planck equation (FPE) describing the
evolution of the velocity distribution. The derivation of the
FPE for two-photon cooling follows
the standard lines that can be found in the literature (see \cite{ref:FPE}). 
If $|V| \gg 1$ the coefficients in the resulting equation
can be expanded up to second order in $1/|V|$ (this is the so-called 
{\em hypothesis of small jumps}). Moreover, if $K|V| \ll |\bar{\delta}|,g$
the resulting expression can be expanded up to the order $V$. The
resulting FPE reads
\begin{equation}
  {\frac{\partial n}{\partial t}}=2\bar{\Gamma}^{\prime }
  {\frac{\partial (Vn)}{\partial V}}+\left( 2\bar{%
\Gamma}+{\frac{\Gamma _{0}}{2}}\right) {\frac{\partial ^{2}n}{\partial V^{2}}%
}  
\label{eq:FPE}
\end{equation}
where $\bar{\Gamma}\equiv \Gamma_{-1}(0)=\Gamma _{0}/4$ and $\bar{\Gamma}%
^{\prime }$ is the $V$-derivative of $\Gamma_{-1}$ evaluated at $V=0$.
Multiplying this equation by $V^2$ and integrating over $V$ one easily
obtains:
\begin{equation}
  { d \langle V^2 \rangle \over dt } =
  - 4 \bar{ \Gamma ^\prime }  \langle V^2 \rangle +
  \left( 4 \bar{ \Gamma } + \Gamma_0  \right)
  \label{eq:Tevolv}
\end{equation}
As $ \langle V^2 \rangle = T/T_r$, this equation shows that the
characteristic relaxation time is $(4 \bar{\Gamma^\prime})^{-1}
=(g \Gamma _{2P} \bar{I}^2 \bar{\delta} K)/(4\bar{\delta}^2+g^2)$.

\begin{figure}[tbp]
  \begin{center}
    \epsfig{figure=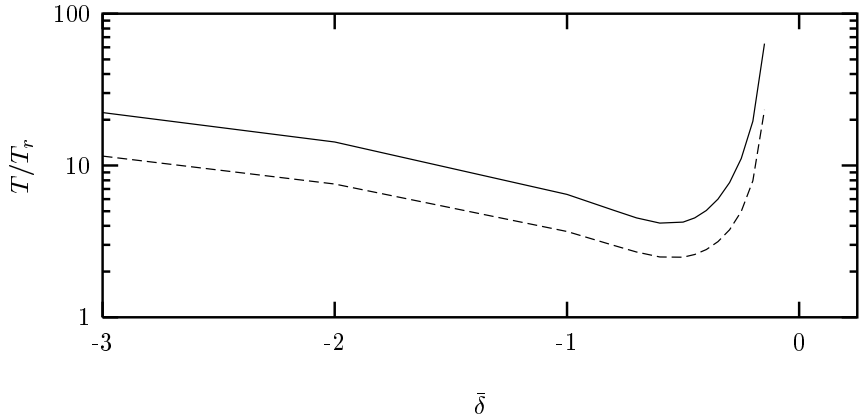,width=15cm}
    \end{center}
  \caption{Dependence of the temperature (log scale) on the detuning.
    The full curve takes into account all two-photon transitions,
    whereas in the dotted curve the Doppler-free transitions have been
    suppressed. The plot shows that the effect of the
latter is to increase the temperature by a factor of two, in agreement
with the FPE prediction.}
\label{fig:Effect_DF}
\end{figure}

The equilibrium temperature is then given by
\begin{equation}
  { T \over T_r } = { 2 \bar{ \Gamma} + \Gamma _0/2 \over
    2 \bar{ \Gamma ^\prime} } 
  = { \bar{ \delta}^2+g^2/4 \over K |\bar{\delta}| }
  \label{eq:Teq}
\end{equation}
This results confirms that the Doppler-free two-photon transitions,
corresponding to the contribution $\Gamma _{0}/2=2\bar{\Gamma}$
in Eq. (\ref{eq:Teq}) increase the equilibrium temperature
(at least in the range
of validity of the FPE) by a factor 2. This fact can also be
verified from the numerical simulations, as shown in Fig.
\ref{fig:Effect_DF}, where the dotted curve corresponds to
the temperature obtained without DF transitions.
As in one-photon Doppler-cooling, the equilibrium 
temperature is independent of the laser intensity (but the time need to
achieve cooling obviously increases as the laser intensity diminishes).

\begin{figure}[tbp]
  \begin{center}
    \epsfig{figure=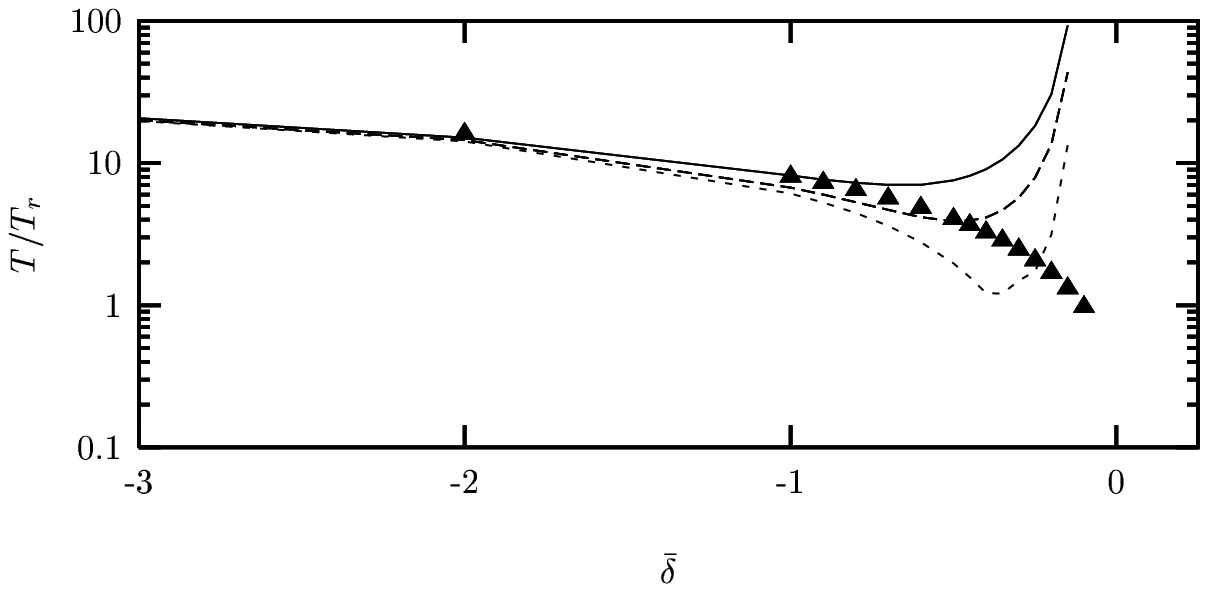,width=15cm}
    \end{center}
\caption{Dependence of the temperature (log scale) on the detuning for three
  values of $g$:  0.9 (full line) 0.5 (dashed line)  and
  0.09 (dotted line). The triangles correspond to the calculation
  based on Eq.(\protect \ref{eq:Teq}) for $g=0.5$ and show the breaking of the
Fokker-Planck approach at temperatures close to $T_r$. The curve
corresponding to $g=0.09$ shows that the minimum temperature is
very close to the recoil limit.}
\label{fig:Effect_g}
\end{figure}

Note that the range of validity of the FPE is $|V| \gg 1$.
It thus fails when the temperature approaches the
recoil temperature (or, in other words, $|V| \approx 1$).
Fig. \ref{fig:Effect_g} shows the dependence of the equilibrium
temperature as a function of the detuning for different values of
parameter $g$. The minimum temperature is clearly
reduced by the decreasing
of $g$, up  to values close to the recoil temperature $T_r$.
Moreover, the figure 
shows that the minimum temperature
generally agrees with the theoretical predictions: it is governed {\em both}
by the effective linewidth $g$ of the excited state {\em and} by the
detuning, the optimum value being $\bar{ \delta} \approx -g/2$
(in the range of validity of the FPE). A reasonably good
agreement between numerical data and the FPE prediction within its
range of validity is also observed.

Let us finally note that an interesting practical possibility is to change
the quenching parameter as the cooling process proceeds. One starts with a
high value of $g$ in order to rapidly cool the atoms to a few recoil
velocities. Then, the quenching parameter {\em and} the detuning are
progressively decreased, achieving temperatures of order of the recoil
temperature. A detail study of the procedure optimizing
the final temperature is however out of the scope of the present paper.

In conclusion, we have suggested and analyzed, both analytically and
numerically, the using of $1S-2S$ two-photon transition together with the
quenching of the $2S$-state to cool hydrogen atoms to velocities
approaching the recoil limit. The quenching ratio gives an additional,
dynamically controllable parameter.

Laboratoire de Physique des Lasers, Atomes et Mol{\'e}cules (PhLAM) is UMR
8523 du CNRS et de l'Universit{\'e} des Sciences et Technologies de Lille.
Centre d'Etudes et Recherches Lasers et Applications (CERLA) is supported by
Minist\`{e}re de la Recherche, R\'{e}gion Nord-Pas de Calais and Fonds
Europ\'{e}en de D\'{e}veloppement Economique des R\'{e}gions (FEDER).

\vspace{1cm} %FIGURE 1
\vspace{1cm}

%FIGURE 2
\vspace{1cm}

%FIGURE 3
\vspace{1cm}

                                %FIGURE 4

\end{document}